\documentclass[%
 reprint,
superscriptaddress,
 amsmath,amssymb,
 aps,
float]{revtex4-1}

\usepackage{subfiles}

\usepackage{graphicx}% Include figure files
\usepackage{dcolumn}% Align table columns on decimal point
\usepackage{bm}% bold math
\usepackage{amsmath}
\usepackage{physics}
\usepackage{gensymb}
\usepackage{placeins}
\makeatletter
\def\maketitle{
\@author@finish
\title@column\titleblock@produce
\suppressfloats[t]}
\makeatother

\begin{document}
\title{High-Stability Single-Ion Clock with $5.5\times10^{-19}$ Systematic Uncertainty}
\author{Mason C. Marshall}
\email{mason.marshall@nist.gov}
\affiliation{Time and Frequency Division, National Institute of Standards and Technology, Boulder, CO, USA}
\author{Daniel A. Rodriguez Castillo}
\affiliation{Time and Frequency Division, National Institute of Standards and Technology, Boulder, CO, USA}
\affiliation{Department of Physics, University of Colorado, Boulder, CO, USA}
\author{Willa J. Arthur-Dworschack}
\affiliation{Time and Frequency Division, National Institute of Standards and Technology, Boulder, CO, USA}
\affiliation{Department of Physics, University of Colorado, Boulder, CO, USA}
\author{Alexander Aeppli}
\affiliation{Department of Physics, University of Colorado, Boulder, CO, USA}
\affiliation{JILA, National Institute of Standards and Technology and the University of Colorado, Boulder, CO, USA}
\author{Kyungtae Kim}
\affiliation{Department of Physics, University of Colorado, Boulder, CO, USA}
\affiliation{JILA, National Institute of Standards and Technology and the University of Colorado, Boulder, CO, USA}
\author{Dahyeon Lee}
\affiliation{Department of Physics, University of Colorado, Boulder, CO, USA}
\affiliation{JILA, National Institute of Standards and Technology and the University of Colorado, Boulder, CO, USA}
\author{William Warfield}
\affiliation{Department of Physics, University of Colorado, Boulder, CO, USA}
\affiliation{JILA, National Institute of Standards and Technology and the University of Colorado, Boulder, CO, USA}
\author{J. Hinrichs}
\affiliation{Time and Frequency Division, National Institute of Standards and Technology, Boulder, CO, USA}
\affiliation{Institute of Quantum Optics, Leibniz University Hannover, Hannover, Germany}
\author{Nicholas V. Nardelli}
\affiliation{Time and Frequency Division, National Institute of Standards and Technology, Boulder, CO, USA}
\author{Tara M. Fortier}
\affiliation{Time and Frequency Division, National Institute of Standards and Technology, Boulder, CO, USA}
\author{Jun Ye}
\affiliation{Department of Physics, University of Colorado, Boulder, CO, USA}
\affiliation{JILA, National Institute of Standards and Technology and the University of Colorado, Boulder, CO, USA}
\author{David R. Leibrandt}
\affiliation{Time and Frequency Division, National Institute of Standards and Technology, Boulder, CO, USA}
\affiliation{Department of Physics, University of Colorado, Boulder, CO, USA}
\affiliation{Department of Physics and Astronomy, University of California, Los Angeles, CA, USA}
\author{David B. Hume}
\email{david.hume@nist.gov}
\affiliation{Time and Frequency Division, National Institute of Standards and Technology, Boulder, CO, USA}
\affiliation{Department of Physics, University of Colorado, Boulder, CO, USA}

\date{Draft: \today}

\begin{abstract}
We report a single-ion optical atomic clock with fractional frequency uncertainty of $5.5\times10^{-19}$ and fractional frequency stability of $3.5 \times10^{-16}/\sqrt{\tau/\mathrm{s}}$, based on quantum logic spectroscopy of a single $^{27}$Al$^+$ ion.  A co-trapped $^{25}$Mg$^+$ ion provides sympathetic cooling and quantum logic readout of the $^{27}$Al$^+$ $^1$S$_0\leftrightarrow^3$P$_0$ clock transition.  A Rabi probe duration of 1 s, enabled by laser stability transfer from a remote cryogenic silicon cavity across a 3.6 km fiber link, results in a threefold reduction in instability compared to previous $^{27}$Al$^+$ clocks.  Systematic uncertainties are lower due to an improved ion trap electrical design, which reduces excess micromotion, and a new vacuum system, which reduces collisional shifts.  We also perform a direction-sensitive measurement of the ac magnetic field due to the RF ion trap, eliminating systematic uncertainty due to field orientation.
\end{abstract}

\maketitle
\newcommand{\w}{3.5in}

\newcommand{\DopplerMain}{
\begin{figure}[tbp!]
\begin{center}
\includegraphics*[width=\columnwidth]{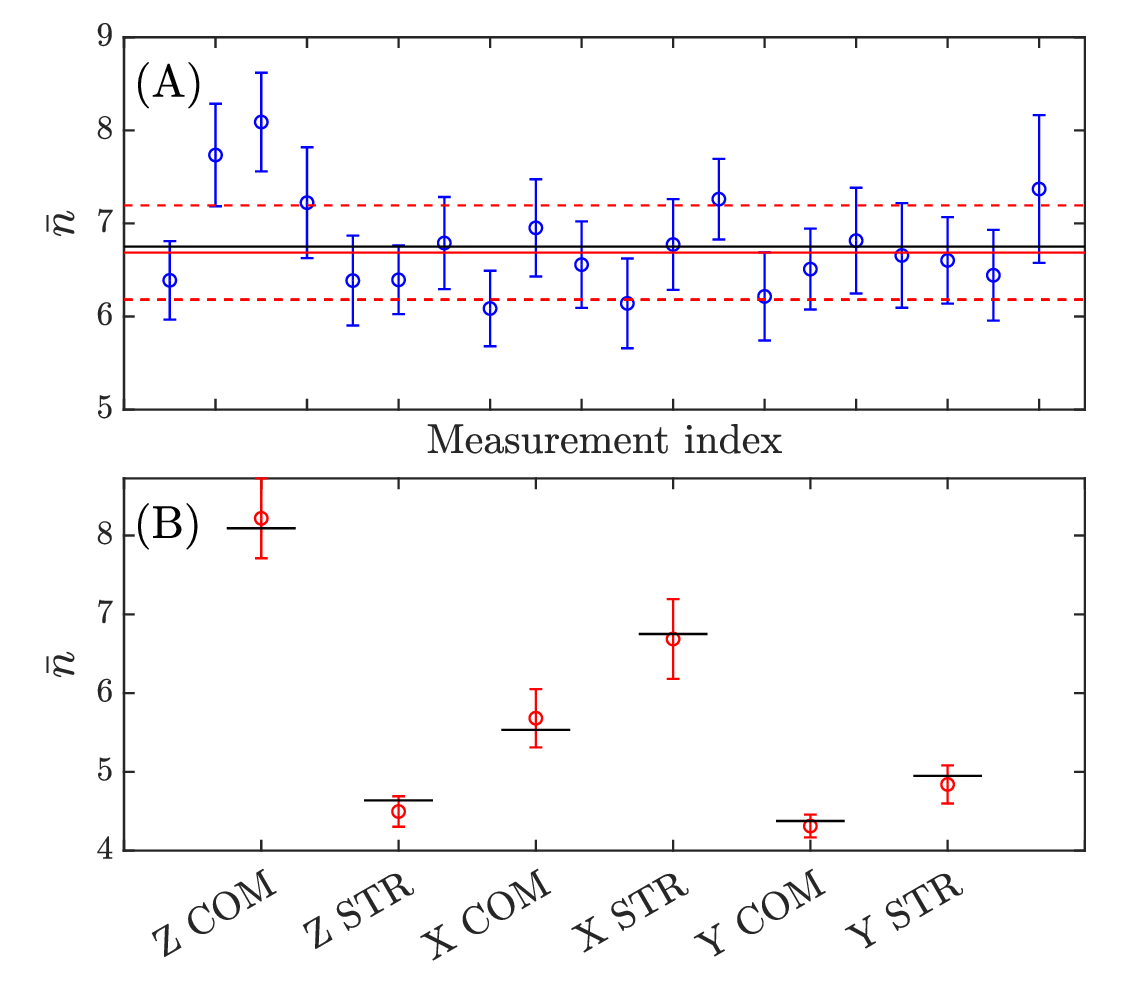}
\caption{\textbf{(A)} Repeated measurements of the motional quantum number for one of the radial stretch modes, performed over several months.  Error bars are 1-$\sigma$ uncertainty.  Red solid and dashed lines are the weighted mean and standard deviation of the measurements.  Black line is the calculated Doppler limit for our nominal trap and laser parameters.  \textbf{(B)} Weighted means and standard deviations for all six modes. 
 Z modes are along the trap axis; X and Y modes are radial.  STR: stretch; COM: center-of-mass.  Black lines are the calculated Doppler limit for each mode.}
\label{fig:DopplerMain}
\end{center}
\end{figure}
}

\newcommand{\FigureOne}{
\begin{figure}[tbp!]
\begin{center}
\includegraphics*[width=\columnwidth]{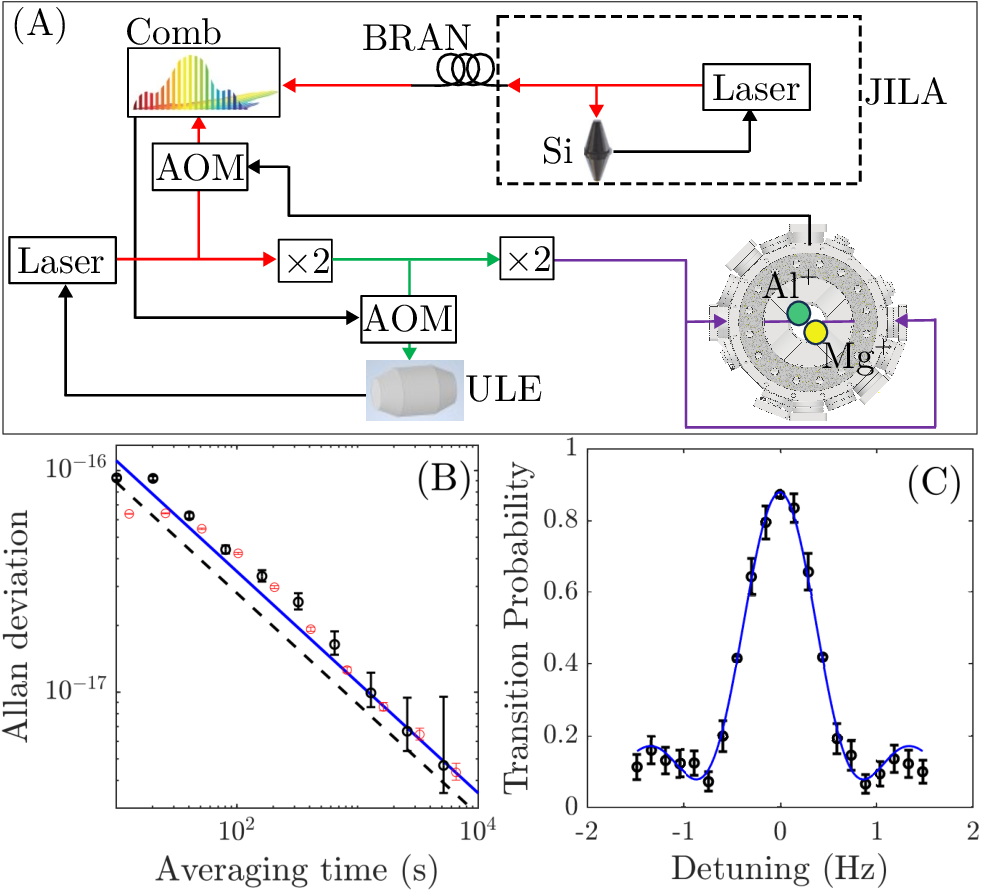}
\caption{\textbf{(A)} Schematic of the optical network for laser stability transfer between the JILA silicon cavity and the Al$^+$ clock laser.  See text for details.  Black lines represent electronic feedback, all others represent path-length stabilized laser beam paths.  BRAN: Boulder Research and Administrative Network (3.6 km fiber link); AOM: Acousto-optic modulator used for frequency stabilization; Si: cryogenic silicon cavity; $\times$2: frequency doubling stage; ULE: room-temperature ultralow-expansion glass cavity.  \textbf{(B)} Overlapping Allan deviation of the frequency ratio $\nu_{\mathrm{Al}^+}/\nu_{\mathrm{Sr}}$ (black points).  Asymptotic fit (blue line) gives a fractional frequency stability of $3.5\times 10^{-16}/\sqrt{\tau / \mathrm{s}}$ beyond the servo time of $\sim80$ s (where $\tau$ is the averaging time).  Dashed line represents the projection noise limit.  Red points are overlapping Allan deviation of a simulation incorporating experimental noise and duty cycle.  Error bars are 68 \% confidence interval of Allan deviation.  \textbf{(C)} $^{27}\mathrm{Al}^+$ clock transition lineshape with a 1 s probe.  Error bars are 68 \% binomial confidence interval of transition probability.}
\label{fig:Figure1}
\end{center}
\end{figure}
}

\newcommand{\CoolingStark}{
\begin{figure}[tbp!]
\begin{center}
\includegraphics*[width=\columnwidth]{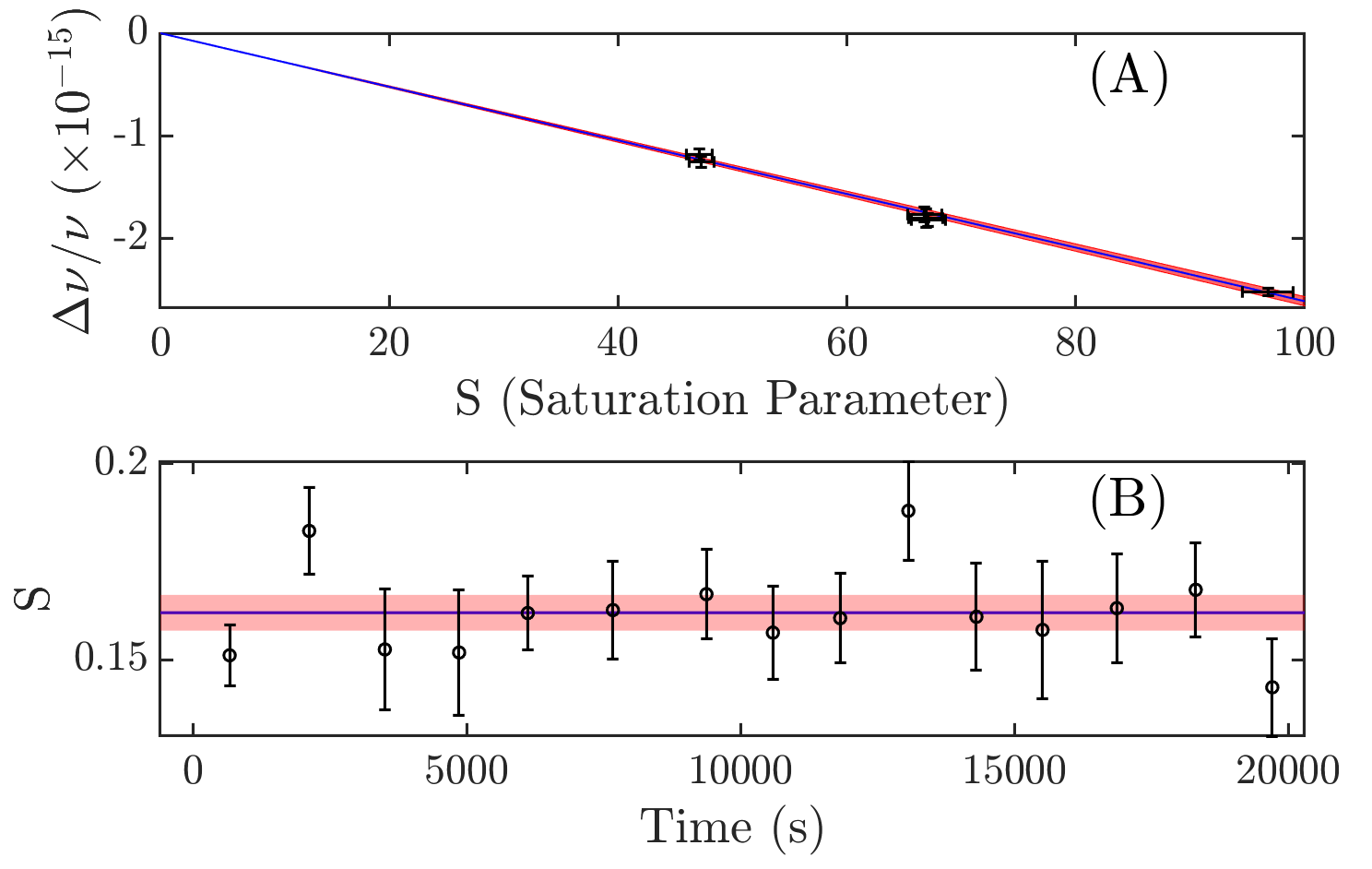}
\caption{Stark shift due to cooling light.  \textbf{(A)} Six measurements of the frequency shift on the clock transition were performed at three different Stark shifting beam powers.  Error bars are 1-$\sigma$ uncertainty.  Blue line and red shaded region give the mean shift per saturation parameter and its weighted standard deviation.  \textbf{(B)} Time series of saturation parameters measured during a day of clock operation, showing no evidence of long-term drift.  Error bars are 1-$\sigma$.  Blue line and red shaded region are the saturation parameter and its 1-$\sigma$ uncertainty, S = 0.1621 $\pm$ 0.0046, extracted from fitting the full day of data.}
\label{fig:CoolingStark}
\end{center}
\end{figure}
}

\newcommand{\BACfig}{
\begin{figure}[tbp!]
\begin{center}
\includegraphics*[width=\columnwidth]{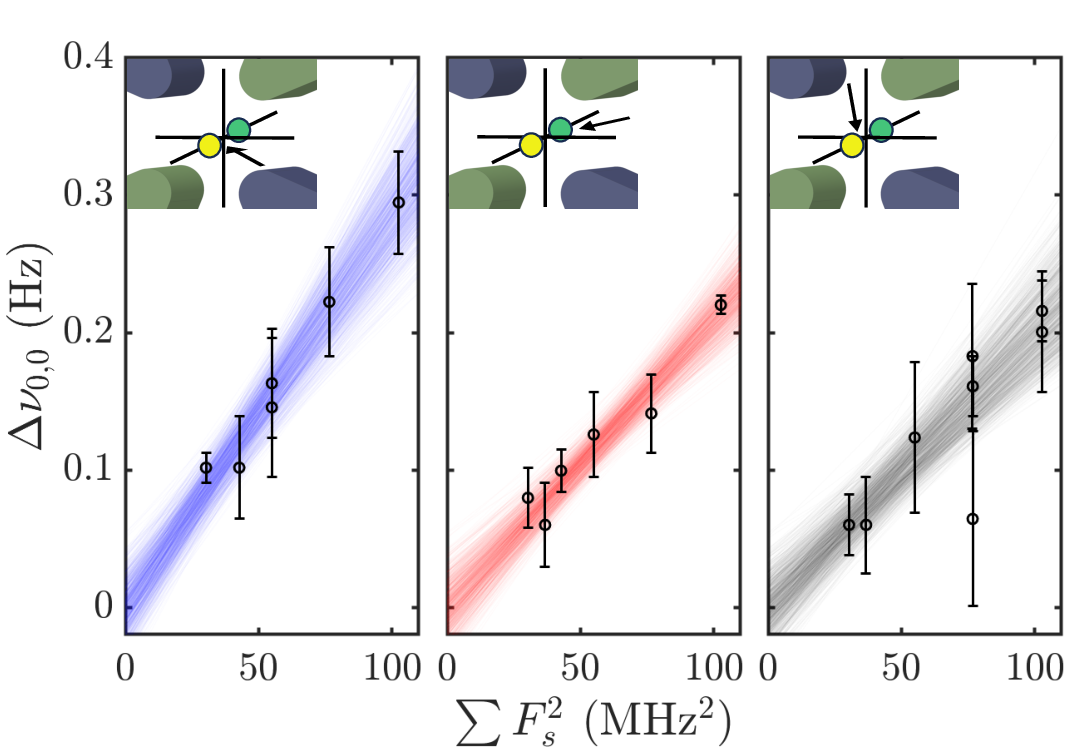}
\caption{Frequency shift on the $^{25}$Mg$^+$ ${|F=3, m_F=0\rangle\leftrightarrow|F=2, m_F = 0\rangle}$ ground state hyperfine transition due to the trap-induced ac magnetic field.  This shift is linear with $\langle B_{AC}^2\rangle$ and here plotted as a function of trap RF power, represented by the sum of squared trap secular frequencies $\Sigma F_s^2$.  Error bars are 1-$\sigma$ uncertainty.  The quantization field and detection laser are oriented along three nearly orthogonal axes; directions are nearly vertical or at 45 degrees to the trap axis. These three axes are depicted as arrows in each inset, together with trap RF electrodes.  Also shown are a subset of fit results from parametric bootstrapping of the measurement data and the angle of the near-orthogonal axes; see text and \cite{SM}.}
\label{fig:BACfig}
\end{center}
\end{figure}
}

\newcommand{\ErrorBudget}{
\begin{table}[htbp!]
\caption{Fractional frequency shifts and uncertainties for the NIST $^{27}$Al$^+$ quantum logic clock}
\begin{tabular}{lcc}
\hline \hline
Effect & Shift ($10^{-19}$) & Uncertainty ($10^{-19}$) \\ \hline
  Secular motion     &           -114.6         &           3.8         \\ 
  dc quad. Zeeman & -6317.9 & 2.5 \\
  Cooling laser Stark & -37.2 & 2.0 \\
  Blackbody radiation & -30.7 & 1.7 \\
  Excess micromotion & -1.6 & 1.6 \\
  Clock laser Stark & 0 & 0.8 \\
  Background gas collisions & -0.3 & 0.4 \\
  ac quad. Zeeman & -0.54 & 0.06 \\
  First-order Doppler & 0 & $<$1 \\
  AOM phase chirp & 0 & $<$1 \\
  Electric quadrupole & 0 & $<$1 \\
  \\
  Total & -6502.8 & 5.5 \\ \hline \hline
\end{tabular}
\end{table}
}

\label{sec:introduction}

\textit{Introduction} -- Optical atomic clocks based on spectroscopy of dipole-forbidden electronic transitions in isolated, trapped atoms are among the most precise instruments developed, capable of measuring time more precisely than the cesium clocks that currently define the second \cite{LudlowOpticalClockReview2015}.  Accordingly, optical clock frequency ratios are some of the most accurate measurements \cite{BeloyBacon2021,HausserIndiumMultiIon2025}, and are used as probes for new physics \cite{SafronovaNewPhysicsReview2018}, including time variation of fundamental constants \cite{SafronovaFundamentalConstantVariation2019,SherrillFundamentalConstantVariation2023}; violations of local position invariance \cite{LangePositionInvariance2021}; constraints on dark matter \cite{BeloyBacon2021,KennedyDarkMatterClock2020,FilzingerClocksDM2023}; and general relativity at small scales \cite{ChouAlClockRelativity2010,BothwelStrontiumEnsembleRelativity2022}.    Optical clocks based on single trapped ions \cite{BrewerAccuracy2019,ZhiqiangLutetiumComparison2023} and neutral atoms in optical lattices \cite{AppeliStrontiumAccuracy024} have reached fractional frequency uncertainties below  $10^{-18}$; further advances open new possibilities for these investigations.  Additionally, as the scientific community moves towards the redefinition of the second, advances in the state of the art for clock accuracy and stability are critical \cite{DimarcqSecondRoadmap2024}.

The exquisite degree of control and access to environmentally insensitive transitions offered by trapped atomic ions have made them a leading technology for measurement accuracy.  In particular, the $^1$S$_0\leftrightarrow^3$P$_0$ transition in singly-ionized aluminum offers a high transition frequency, long excited-state lifetime, and one of the lowest known sensitivities to blackbody radiation \cite{SafronovaBBRclocks2012}.  In this letter, we report the accuracy and stability evaluation of the current-generation NIST $^{27}$Al$^+$ quantum logic clock.  This clock realizes the lowest fractional frequency uncertainty of any clock to date, at $\Delta\nu/\nu=5.5\times10^{-19}$.  Its fractional instability of $3.5\times10^{-16}/\sqrt{\tau/\mathrm{s}}$ represents a threefold reduction in instability compared to the previous NIST quantum logic clock \cite{BrewerAccuracy2019}.  Critical to these achievements are a more stable clock laser, an improved Paul trap electrical design with reduced excess micromotion, and a $150\times$ improvement in background gas pressure from a new ultrahigh vacuum system.

\label{sec:operation}

\textit{Clock operation and stability} -- The operation of the clock is similar to that described in \cite{ChouComparison2010,BrewerAccuracy2019}.  The clock cycle begins with preparation of the $^{27}$Al$^+$ clock ion into one of the $\lvert{}^1$S$_{0}$, m$_F$=$\pm$5/2$\rangle$ states via optical pumping on the $^1$S$_0\leftrightarrow^3$P$_1$ transition.  Sympathetic cooling on the $^{25}$Mg$^+$ logic ion then brings the ion pair to the Doppler temperature limit.  Finally, we probe the $^{27}$Al$^+$ clock transition using Rabi spectroscopy followed by quantum logic readout \cite{SchmidtQLS2005,HumeQLS2007}.  

We probe both the $\mathrm{m_F}=+5/2$ and $\mathrm{m_F}=-5/2$ $^1\mathrm{S}_0\leftrightarrow^3\mathrm{P}_0$ transitions, and generate a ``virtual'' first-order magnetic-field insensitive transition from their mean frequency \cite{RosenbandAlClockObservation2007}.  Additionally, we alternate probing from opposite directions, with the two probe beams counterpropagating through the same single-mode optical fibers; the average of opposite directions is insensitive to possible first-order Doppler shifts due to ion motion.  Both probe laser acousto-optic modulators (AOMs) are switched on during each cycle, with the inactive direction detuned by 100 kHz.  This maintains a nearly identical electric- and magnetic-field environment for probes in either direction and either side of the $^{27}$Al$^+$ Zeeman structure.

Decoherence of the $^{27}$Al$^+$ clock laser ($\lambda_L\simeq267$ nm) limited previous generations to a 150 ms probe time.  We extend the laser coherence by transferring the stability of the JILA cryogenic silicon cavity \cite{MateiSiCavity2016} to the clock laser.  An ultrastable laser locked to this cavity travels over a 3.6 km path-length-stabilized fiber link to NIST.  This fiber link via the Boulder Research and Administrative Network has been described previously \cite{BeloyBacon2021}; for this work, it has been modified as shown in Fig. \ref{fig:Figure1}A, such that the cavity-stabilized light serves as reference for an Er:Yb frequency comb \cite{FortierCombReview2019,NardelliErYbComb2023}.  We first pre-stabilize the $^{27}$Al$^+$ clock laser with a Pound-Drever-Hall lock to a room-temperature ultra-low expansion glass cavity.  With a lower bandwidth of $\sim$10 kHz, we then modulate the drive frequency of an AOM to stabilize the beatnote between clock laser and frequency comb.  With this improved clock laser stabilization, we can operate the clock using a 1 s probe duration without substantial loss of atom-laser coherence.

\FigureOne

To operate with a 1 s probe, we control effects which scale with the probe time.  Previous $^{27}$Al$^+$ clocks have been operated by ground-state cooling before the clock probe \cite{ChenCooling2017}; however, with a 1 s probe time, the heating rate in our system would yield average motional phonon occupation numbers above the Doppler limit.  We therefore continuously Doppler cool the logic ion during the clock probe and carefully characterize the temperature at the Doppler limit.  

Since Doppler cooling produces a thermal distribution \cite{StenholmLaserCoolingreview1986,WinelandLaserCoolingLimits1987}, we expect the main nonthermal component to arise from collisions with background gas, which are accounted for in the collision shift \cite{HankinBackgroundgas2019}.  The probability of a collision between the trapped ions and background gas scales linearly with the probe time.  To reduce the collision rate, we developed an all-titanium vacuum system to minimize hydrogen outgassing, as well as designing for improved conductance and hydrogen pumping speed.  These improvements yielded a base pressure of $(2.5 \pm 1.3) \times 10^{-10}$ Pa $[(1.8\pm0.9)\times10^{-12}$ $\mathrm{Torr}]$ as measured by the reorder rate of a two-ion crystal \cite{SM}, sufficient to suppress the collision effect for long probe times.  This also reduces the rate of aluminum hydride formation, decreasing the need to stop the clock and reload ions during a measurement run.  

In addition to readout and state preparation, auxiliary operations are interleaved with clock interrogation to stabilize the ion order and measure the intensity of the cooling light.  The clock interrogation has a duty cycle of $\sim80~\%$.  The clock-laser path length is stabilized through all optical fibers from the frequency comb to the trap vacuum chamber.  An active magnetic field servo with $\sim$1 kHz bandwidth adjusts shim coils surrounding the optical table based on fluxgate sensors located on opposite sides of the vacuum chamber; this stabilizes the quantization field and minimizes noise at 60 Hz and harmonics.

We measure the $^{27}$Al$^+$ clock stability by comparison to the JILA Sr optical lattice clock \cite{AeppliJILASrClock2024}, with results shown in Fig. \ref{fig:Figure1}B.  The optical lattice clock fractional stability is $<1\times10^{-16}/\sqrt{\tau/\mathrm{s}}$, meaning the comparison stability of $3.5\times10^{-16}/\sqrt{\tau/\mathrm{s}}$ is dominated by the $^{27}$Al$^+$ clock.  To our knowledge, this represents the lowest instability of any ion clock reported to date.  In the future, this stability could be improved by further extending the probe time using differential spectroscopy \cite{KimDiffspec2023}, or by extending the clock to multiple spectroscopy ions \cite{CuiScalableQLS2022, KellerMultIonConf2024,PelzerMultiIonCa2024}.

In the following paragraphs, we describe the evaluation of systematic shifts to the $^{27}$Al$^+$ clock transition.

\ErrorBudget

\newpage

\textit{Secular motion} -- The relativistic time-dilation shift, or second-order Doppler shift, due to secular motion in the ion trap \cite{WinelandLaserCoolingLimits1987} is the largest source of systematic uncertainty in this clock.  We evaluate the energy in the secular modes of the two-ion crystal using sideband thermometry \cite{LeibfriedQuantumDynamics2003,WinelandExperimentalIssues1998}. To quantify the statistical uncertainty and test the repeatability of sideband thermometry at the Doppler limit, we repeat the measurements many times over the course of several months.  The results are consistent within the measurement uncertainty with the calculated Doppler limit for our nominal laser geometry and secular mode frequencies.  Figure \ref{fig:DopplerMain} shows the record of measurements for one of the radial ``stretch'' modes, with the others shown in supplemental material \cite{SM}.

\DopplerMain

We controlled for potential systematic shifts in this temperature measurement by varying sideband pulse duration, altering the motional frequency spectrum and using higher-order sidebands for the analysis.  We found no significant dependence of the results on any of these changes.  We take the weighted standard deviation of all measurements for each secular mode to be the uncertainty on that mode's temperature.  This accounts for possible instability in the underlying value as well as statistical uncertainty in the individual measurements.  Since the uncertainties for each mode are statistical, we add them in quadrature to get a total secular motion shift of $\Delta\nu/\nu=-(114.6\pm 3.8) \times 10^{-19}$.

\textit{Quadratic Zeeman shift} -- While the first-order Zeeman shift is eliminated by averaging transitions with opposite magnetic-field dependence, there remains a quadratic Zeeman shift.  We write this as $\Delta\nu/\nu=C_2\langle B^2\rangle$, where $C_2$ is the quadratic Zeeman coefficient \cite{BrewerMagconst2019} and $\langle B^2\rangle=\langle B_{DC}\rangle^2+\langle B_{AC}^2\rangle$.  The frequency difference between Zeeman levels gives a real-time measure of the quantization field $B_{DC}$, while $B_{AC}$ is measured via hyperfine spectroscopy on the $^{25}$Mg$^+$ ion \cite{SM}.  The average dc quadratic Zeeman shift is $\Delta\nu/\nu=-(6317.9\pm 2.5)\times10^{-19}$, where the exact value for any given day of operation depends on the measured $B_{DC}$, but the uncertainty is not affected at the given level of precision.  Compared to the previous-generation $^{27}$Al$^+$ clock, we operate at a lower magnetic field of 0.10 mT, which reduces the amplitude and uncertainty of the dc quadratic Zeeman shift. 

\BACfig

$\langle B_{AC}^2\rangle$ is dominated by the magnetic field due to the trap RF drive at $\Omega/2\pi =70.86$ MHz.  We measure $\langle B_{AC}^2\rangle$ by observing a frequency shift on the first-order field-insensitive transition ${|F=3, m_F=0\rangle\leftrightarrow|F=2, m_F = 0\rangle}$ in the $^{25}$Mg$^+$ ground state as a function of RF drive power, while using the first-order sensitive transition ${|F=3, m_F=-3\rangle\leftrightarrow|F=2, m_F = -2\rangle}$ to subtract $\langle B_{DC}\rangle^2$ \cite{BrewerMagconst2019}.  This measurement is sensitive to the direction of the ac magnetic field relative to the quantization axis \cite{BarrettOscillating2018}.  To remove this directional ambiguity, we repeat the measurement with the quantization axis oriented in three nearly orthogonal directions.  We use three pairs of magnetic-field coils to change the magnetic field direction, and propagate the 280 nm laser beams used for $^{27}$Mg$^+$ state preparation and readout along each quantization axis.  In each of these three conditions, we measure the hyperfine frequency shift at a range of different trap RF drive powers.  Simultaneously fitting the results gives the ac quadratic Zeeman shift without directional ambiguity, where we quantify the uncertainty using a parametric bootstrapping method~\cite{EfronBootstrap1979}.  At our operating condition, this is $\langle B_{AC}^2\rangle=0.85\pm0.09$ $\mu \mathrm{T}^2$; this corresponds to a clock frequency shift of $\Delta \nu/\nu=-(0.54\pm0.06)\times10^{-19}$.  Fig. \ref{fig:BACfig} shows the measured frequency shifts, as well as a subset of the simultaneous fit results projected back onto the three measurement axes.  See supplemental material \cite{SM} for more details.

\textit{Cooling laser Stark shift} -- The 280 nm laser beam used for continuous Doppler cooling of the logic ion also illuminates the $^{27}$Al$^+$ spectroscopy ion, inducing an ac Stark shift.  We evaluate this shift by directly measuring the Stark shift on the clock transition induced by a stronger 280 nm beam, then extrapolating to the lower power we use for Doppler cooling.  Specifically, we measure the Stark shift on the clock transition from one of our 280 nm $^{25}$Mg$^+$ Raman beams, typically used for resolved-sideband cooling and quantum logic operations.  While the Raman and Doppler beams are detuned from one another by $\sim$50 GHz, this is small compared to the detuning from the deep UV transitions that dominate the clock transition differential polarizability \cite{RosenbandBBR2006}; the difference in Stark shift is therefore negligible.

The Rabi rates of Raman transitions, together with power ratios between beams in a Raman pair, give an absolute calibration of the saturation parameter $S$ of the Stark shifting beam; see supplemental material \cite{SM} for more details.  We measure at several different Stark-shifting beam powers, as shown in Fig. \ref{fig:CoolingStark}A, obtaining a shift of $\Delta\nu/\nu=-(2.59\pm0.06)\times10^{-17}\times S$.

\CoolingStark

During clock operation, we regularly measure the depumping rate of the $^{25}$Mg$^+$ qubit dark state to determine the saturation parameter.  We apply a correction based on the average measurement across the day and the calibrated ratio of laser powers at the two ions' positions.  Fig. \ref{fig:CoolingStark}B shows measurements across a typical day of operation, binned into approximately 20 minute intervals.  In the absence of any long-term drift, we take the full day's worth of saturation parameter measurements and apply an average correction to the day's clock data.  Repeating this procedure across many days of clock operation, we obtain an average shift and uncertainty of $\Delta\nu/\nu=-(37.2\pm 2.0) \times 10^{-19}$, where the exact value for any given day of operation depends on the measured saturation parameter.

\textit{Blackbody radiation} -- The $^{27}$Al$^+$ clock transition has one of the lowest polarizabilities due to blackbody radiation of any existing atomic clock \cite{SafronovaBBRclocks2012}.  This was recently measured \cite{WeiPolarizability2024} with high precision in a $^{27}$Al$^+$-$^{40}$Ca$^+$ clock, using the Stark shift on the $^{40}$Ca$^+$ clock transition as a reference for the intensity of a Stark shifting infrared laser and giving a polarizability of $(6.86\pm 0.23)\times10^{-42}$ $\mathrm{Jm}^2/\mathrm{V}^2$.  We measure the temperature of our apparatus using a set of six thermocouples - three located on the trap and support structure, and three on the surrounding vacuum chamber - constraining the ion's blackbody environment to be $24.0\pm 3.3$ $\degree$C, for a Stark shift due to blackbody radiation of $\Delta\nu/\nu=-(30.7 \pm 1.7) \times 10^{-19}$.

\textit{Excess micromotion} -- Excess micromotion (EMM) was the largest systematic uncertainty in the previous NIST $^{27}$Al$^+$ clock \cite{BrewerAccuracy2019}.  This is substantially reduced due to an improved Paul trap electrical design, featuring a thicker diamond wafer, thicker sputtered gold traces, and rerouting of the traces to balance capacitances on opposite RF electrodes.  We optimize micromotion compensation voltages at the start of each day of clock operation, and measure again at the end of each day.  These measurements constrain the EMM-induced second-order Doppler shift on the clock transition to $\Delta\nu/\nu=(-1.6\pm 1.6)\times 10^{-19}$; see supplemental material \cite{SM} for more details on trap design and EMM constraints.

\textit{Background gas collisions} -- Frequency shifts due to background gas collisions in a $^{27}$Al$^+$ clock were evaluated in detail in \cite{HankinBackgroundgas2019}, where a strong suppression of Doppler shifts due to collision-induced heating was identified, thanks to the Debye-Waller effect.  The application of continuous Doppler cooling in this work weakens this suppression, as ions which would otherwise not interact with the clock laser are re-cooled during the probe time.  However, the low pressure in our new vacuum system also strongly suppresses this shift.  Accounting for the weaker Debye-Waller suppression, the measured collision rate, and the one-second probe time, we constrain this shift to be $\Delta\nu/\nu=-(0.25\pm0.36)\times10^{-19}$.
See \cite{SM} for additional details.

\textit{First order Doppler shift} -- The previous-generation quantum logic clock observed a first-order Doppler shift of order $5\times10^{-17}$, which was strongly suppressed by averaging clock probes from opposite spatial directions.  In this apparatus, we measure a first order Doppler shift consistent with zero, at $\Delta\nu/\nu=(0.1\pm 1.7)\times 10^{-18}$.  Combined with the suppression from averaging opposite probe directions \cite{BrewerAccuracy2019}, the residual first-order Doppler shift is negligible.  

\textit{Other shifts} -- A possible ac Stark shift due to the clock laser has been investigated, following \cite{BrewerAccuracy2019,ChouComparison2010}.  Because of the lower laser intensity required for a one-second probe, the shift uncertainty is reduced to $\Delta\nu/\nu=\pm0.8\times10^{-19}$.  Possible AOM phase chirp effects and electric quadrupole shifts are also bounded below $10^{-19}$ \cite{BeloyHyperfineShifts2017}.

\textit{Conclusion} -- We have developed a $^{27}\mathrm{Al}^+$ clock with instability of $3.5\times10^{-16}/\sqrt{\tau/\mathrm{s}}$ and inaccuracy of $5.5\times10^{-19}$.  This represents a threefold improvement in stability over previous aluminum ion clocks, while also advancing the state of the art in optical clock accuracy.  The stability could be improved with an even more stable clock laser, enabling longer probe times, or by simultaneous interrogation of a Coulomb crystal including several aluminum ions.  The accuracy is mainly limited by the measurement of the Doppler temperature. We note that a cryogenic system with low heating rates would allow ground state operation with 1 s or longer probe times, while also reducing uncertainty due to background gas collisions and blackbody radiation.  Combined with realistic improvements to the measurement of the quadratic Zeeman coefficient and micromotion amplitude, a clock with accuracy at the $1 \times 10^{-19}$ level is feasible.

\textit{Acknowledgements} --
We thank S. M. Brewer for early design work on the trap and vacuum system; A. Contractor for contributions to the ion fluorescence imaging system; and C.-W. Chou for work on the saturation parameter measurement formalism and for useful discussions.  We thank L. Sonderhouse and S. Scheidegger for careful reading of the manuscript.  We acknowledge support from the Office of Naval Research; the National Institute of Standards and Technology; the National Science Foundation Q-SEnSE Quantum Leap Challenge Institute (grant number OMA-2016244); the V. Bush Fellowship; and the National Science Foundation (award number PHY-2317149).  This Letter is a contribution of the U.S. government, not subject to U.S. copyright.

Data are available from the authors upon reasonable request.

%\reftitle{References}
\bibliographystyle{prsty_gg}
\bibliography{MasonClocksbib.bib}

\clearpage

\title{Supplemental Material for ``High-Stability Single-Ion Clock with $5.5\times10^{-19}$ Systematic Uncertainty''}

\date{Draft: \today}

%\begin{abstract}
%abstract goes here
%\end{abstract}

\maketitle

\renewcommand{\figurename}{Fig.}
\renewcommand{\thefigure}{S\arabic{figure}}

\renewcommand{\tablename}{TABLE}
\renewcommand{\thetable}{S\arabic{table}}

\newcommand{\SuppDoppAll}{
\begin{figure*}[htbp!]
\begin{center}
\includegraphics*[width=\textwidth]{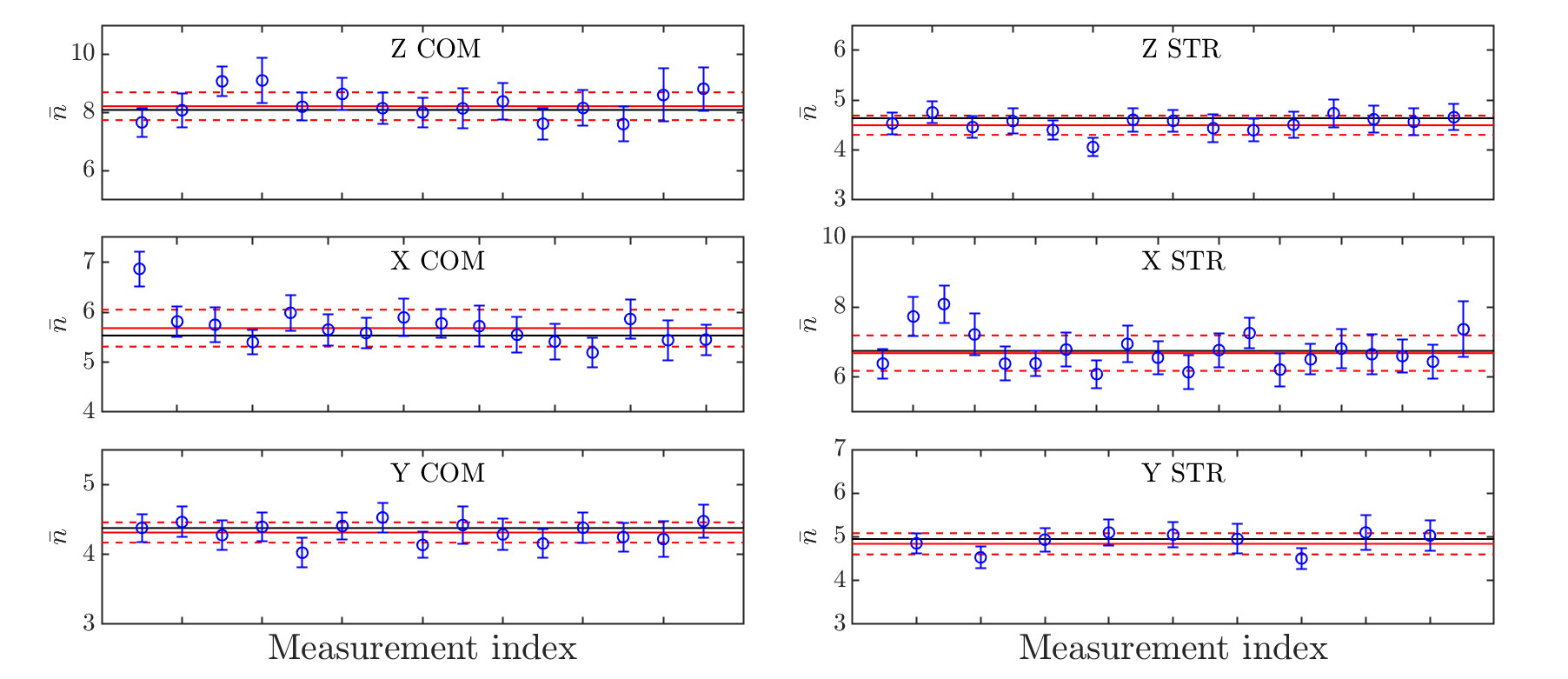}
\caption{Doppler limit secular mode temperature data for all six modes.  Error bars are 1-$\sigma$ uncertainty.  As in Fig. 2 of the main text, the red solid and dashed lines are the weighted mean and standard deviation of the repeated measurements taken over several months, while the black line is the calculated Doppler limit for our nominal trap and laser parameters.}
\label{fig:SuppDoppAll}
\end{center}
\end{figure*}
}

\newcommand{\SuppRaman}{
\begin{figure}[tbp!]
\begin{center}
\includegraphics*[width=\columnwidth]{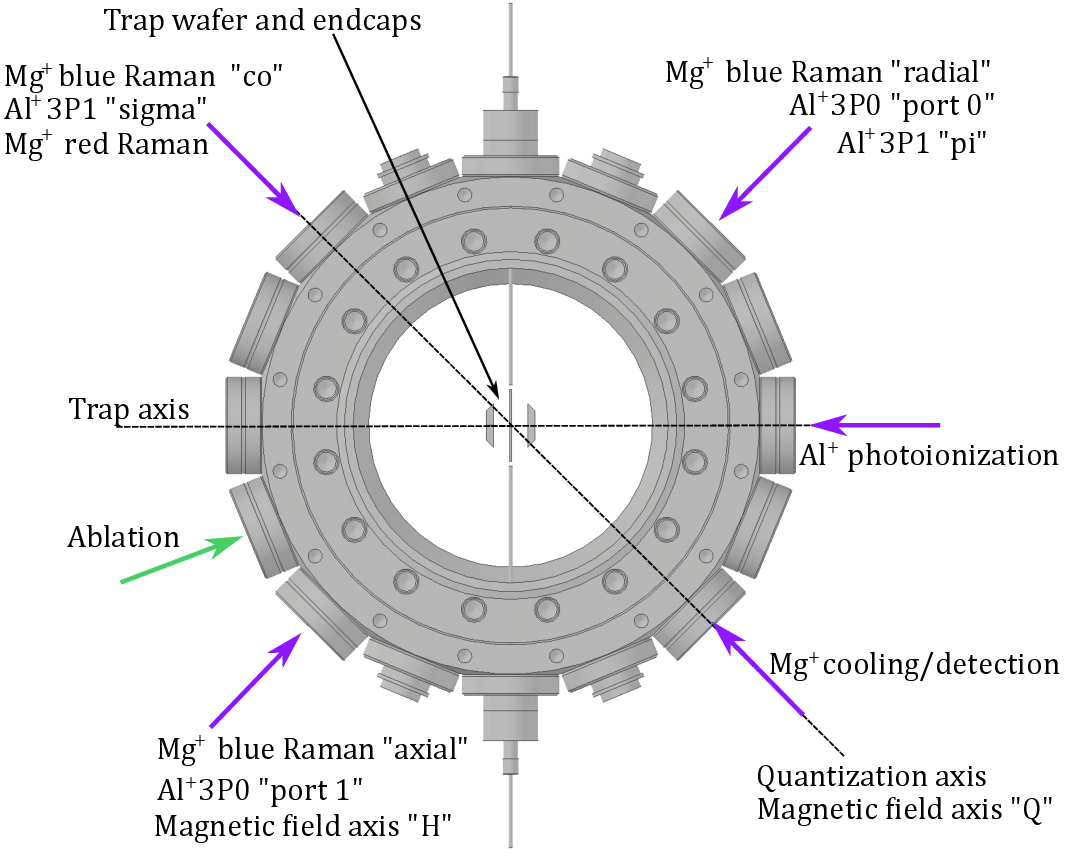}
\caption{Trap and beam geometry.  Not pictured are two beams entering at a $\sim10\degree$ angle from the vertical axis: Al$^+$ $^3\mathrm{P}_1$ ``vertical'' and Mg$^+$ blue Raman ``vertical''.  }
\label{fig:SuppRaman}
\end{center}
\end{figure}
}

\newcommand{\MMSupp}{
\begin{figure}[tbp!]
\begin{center}
\includegraphics*[width=\columnwidth]{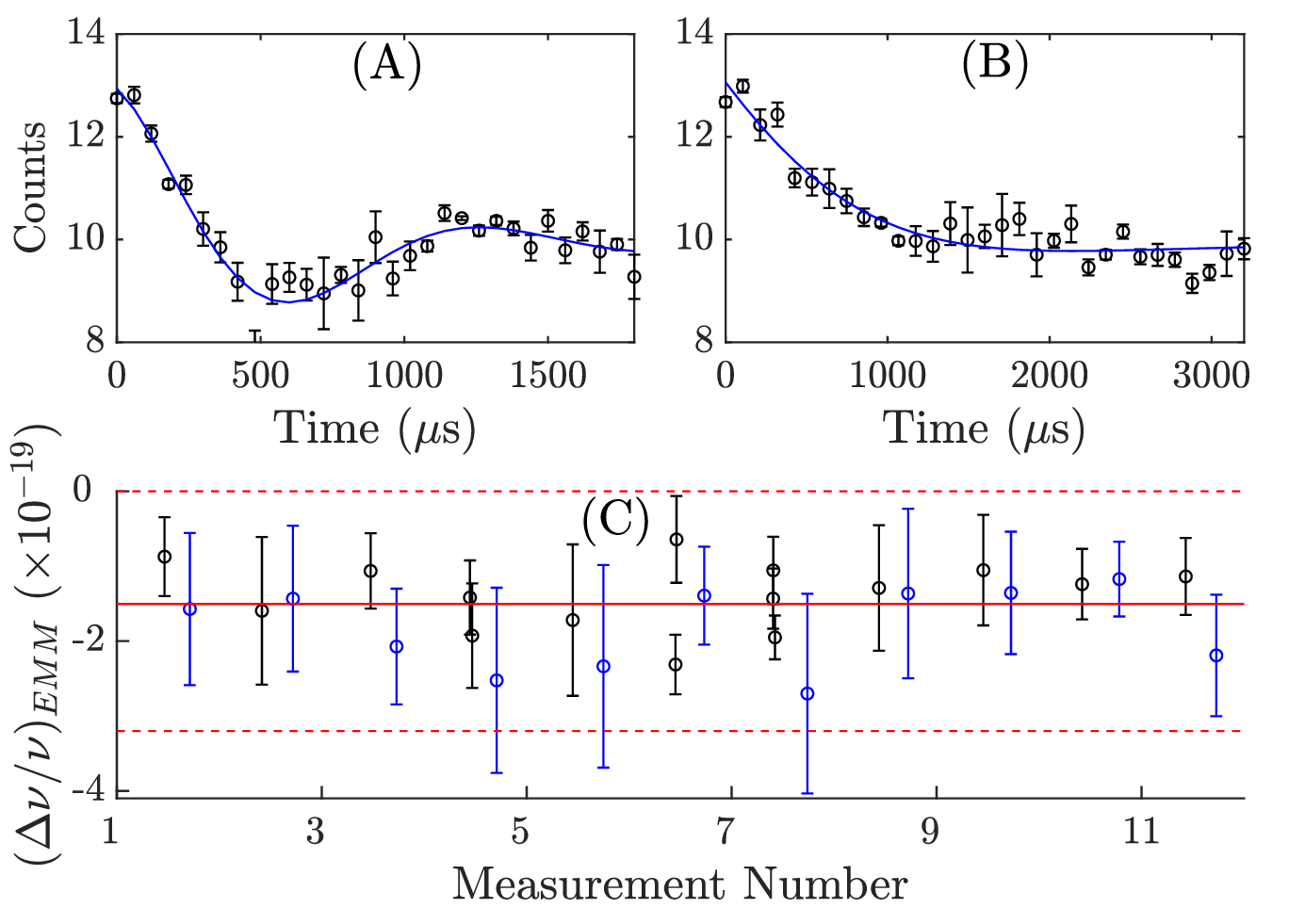}
\caption{Measurements of excess micromotion (EMM).  \textbf{(A)} and \textbf{(B)} Time evolution of micromotion sidebands driven with Raman beam pairs.  (A) shows a measurement with a higher EMM modulation index, resulting in a well-constrained sideband Rabi rate.  (B) shows a measurement where the sideband Rabi rate exceeds the coherence of the transition, placing an upper bound on the EMM rather than a measurement.  Error bars in (A) and (B) are 1-$\sigma$ uncertainty.  \textbf{(C)} EMM shifts derived from measurements taken before (black) and after (blue) several hours of clock operation on different days.  Error bars are 1-$\sigma$ uncertainty.  Red solid and dashed lines represent the weighted mean and the uncertainty assigned to EMM; see text for details.}
\label{fig:MMSupp}
\end{center}
\end{figure}
}

\newcommand{\SuppVacuum}{
\begin{figure}[t]
\begin{center}
\includegraphics*[width=\columnwidth]{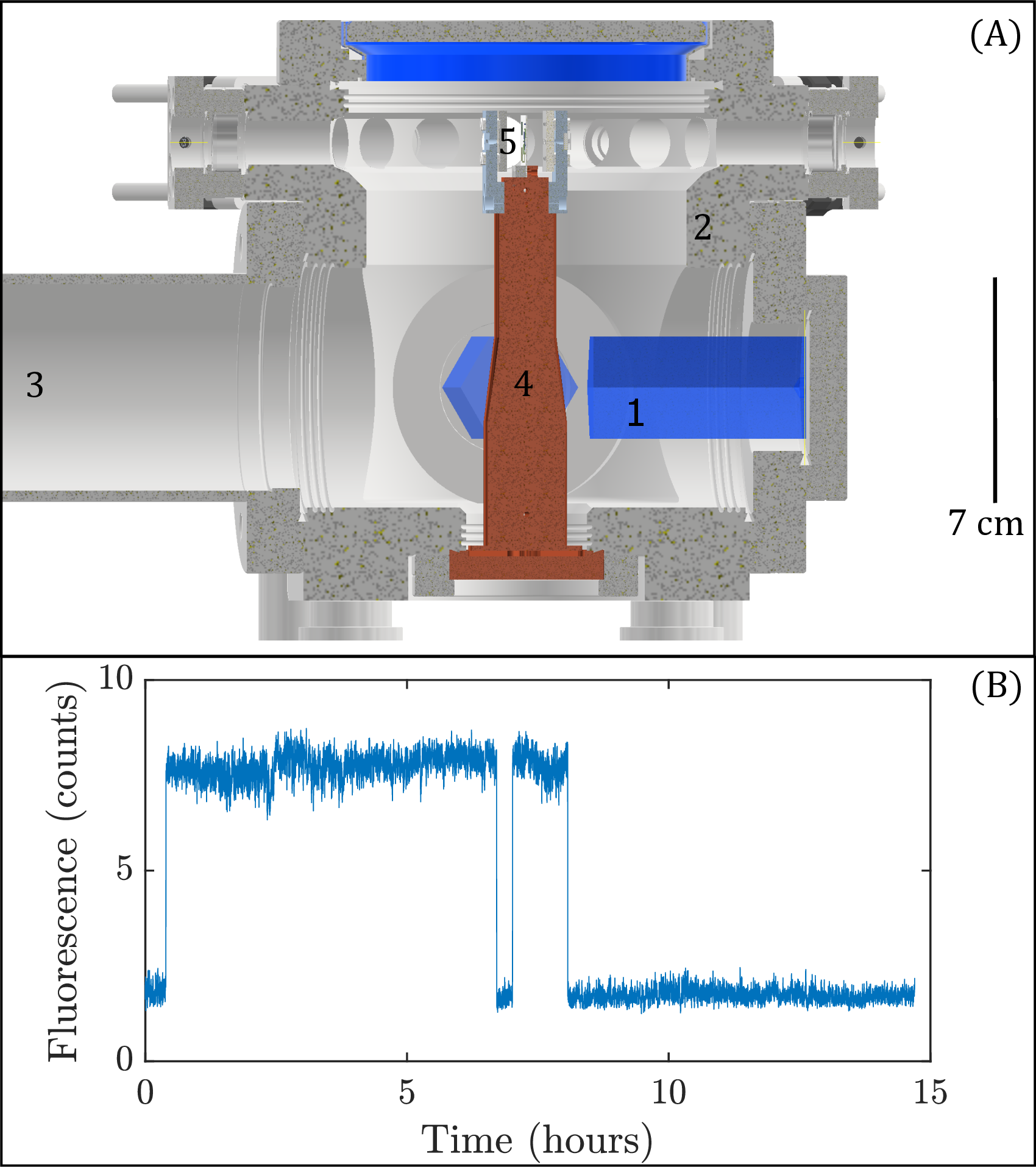}
\caption{\textbf{(A)} Schematic of main vacuum chamber.  Features relevant to vacuum performance are noted as follows.  1: Non-evaporable getter pumps in main chamber have direct line-of-sight to ion trap and high pumping speed for hydrogen.  2: All-titanium construction reduces hydrogen outgassing compared to steel.  3: 2.75" diameter tube provides high conductance to ion pump.  4: Trap support and thermal anchoring post.  5: Ion trap.  \textbf{(B)} Subset of reorder rate data used to determine vacuum pressure.  Ions are addressed with a tightly focused Raman beam.  After the evolution time under the tightly focused beam, the $^{25}\mathrm{Mg}^+$ will fluoresce if the ion pair is in one order and not the other.  Large, fast changes in fluorescence represent reorder events.}
\label{fig:SuppVacuum}
\end{center}
\end{figure}
}

\newcommand{\SecTable}{
\begin{table*}[htbp!]
\caption{\label{tab:SecularMotion}
Secular motion parameters for the $^{25}$Mg$^+$-$^{27}$Al$^+$ crystal.
}
\begin{ruledtabular}
\begin{tabular}{ccccccc}
 & Axial COM & Axial STR & X COM & X STR & Y COM & Y STR \\
\hline
  Frequency [MHz] & 2.16 & 3.75 & 4.22 & 3.48 & 5.37 & 4.75\\ 

  Frequency shift per quantum ($10^{-19}$) &-0.95  & -1.42 & -1.77 & -6.48 & -1.42 & -6.53 \\

  Measured $\bar{n}$ & 8.22 $\pm$ 0.48 & 4.50 $\pm$ 0.19 & 5.68 $\pm$ 0.49 & 6.69 $\pm$ 0.51 & 4.31 $\pm$ 0.14 & 4.84 $\pm$ 0.24\\

Geometric factor $\kappa$ & 1.7 & 1.7 & 2.3 & 2.3 & 2.3 & 2.3 \\
  
Calculated Doppler limit $\bar{n}$ & 8.1 & 4.6 & 5.5 & 6.8 & 4.4 &  5.0\\

\end{tabular}
\end{ruledtabular}
\end{table*}
}

\section{Secular motion}

We measure the secular mode temperature for all six modes using sideband thermometry, with results in Fig. 2b above.  For a thermal state, the occupation probabilities $p_n$ and the average quantum number $\bar{n}$ are related by $\frac{p_{n+1}}{p_n}=\frac{\bar{n}}{\bar{n}+1}$.  For red and blue sideband transition probabilities $P_r$ and $P_b$, this implies $\bar{n}=\frac{P_r}{P_b-P_r}$ \cite{LeibfriedQuantumDynamics2003,WinelandExperimentalIssues1998,TurchetteHeating2000}.  As long as the ion's motion is in a thermal state, this method is valid under a broad range of experimental conditions.  One challenge is that the difference $P_b-P_r$ can be small for $\bar{n}>1$, leading to large statistical fluctuations.  We counteract this by taking sufficiently large data sets and by repeating the measurement over the course of several months to check for reproducibility. Fig. \ref{fig:SuppDoppAll} shows mode temperature data for all six modes.

During mode temperature measurements, we interleave red and blue sidebands to ensure identical conditions.  To eliminate drifts in fluorescence collection efficiency, we also interleave control measurements where no sideband is applied.  We subtract the photons collected after a red or blue sideband from a control measurement to give the red or blue contrast.  
We alternate conditions every 200 measurements, repeating the sequence 200 times in total; each point in Fig. \ref{fig:SuppDoppAll} represents 80,000 total fluorescence measurements.  Error bars in Fig. \ref{fig:SuppDoppAll} are the 1-$\sigma$ uncertainty on the mean of each data set.

\SecTable

We also check for potential systematic errors in this precise measurement of the Doppler-cooled temperature.  For example, the presence of accidental near-degeneracies between motional sidebands of different Zeeman transitions could cause a difference in the apparent Rabi frequency between red and blue sidebands.  To control for this and other possible systematics, we make at least two measurements for each mode in a variety of different conditions.  We modify the radial mode structure by changing the degeneracy-lifting voltage $U_r$ \cite{WinelandExperimentalIssues1998}; change the axial confinement by $\sim$350 kHz; vary the sideband pulse duration between $3\pi/4$ and $5\pi/2$; vary the saturation parameter of the cooling light between 0.11 and 0.18; and change the order of the motional sideband.  After controlling for changes in the predicted Doppler limit at different secular frequencies, none of these measurements give a statistically significant shift from the data presented in Fig. \ref{fig:SuppDoppAll}.  The uncertainty in these parameter-dependence measurements, together with the measured variations in cooling parameters, limit their influence to an additional 3 \% uncertainty.  We add this in quadrature with the statistical uncertainty on each temperature measurement.

We compare these temperatures to the calculated Doppler limit for each mode, using the equation \cite{ItanoCoolingLimits1982}
\begin{equation}
    \bar{n}_{Dopp}=\frac{\gamma}{2\pi f_i}\kappa_i
\end{equation}
where $\gamma$ is the Doppler cooling transition linewidth, the Doppler cooling laser detuning is $\gamma/2$, $f_i$ is the secular frequency for motional mode $i$, and $\kappa_i$ is a geometric factor equal to 1.7 for axial modes and 2.3 for radial modes.  We find that the measurements agree within their uncertainty with the calculated Doppler limit.  Figure \ref{fig:SuppDoppAll} shows the measurement set for each of the six secular modes, together with the weighted mean and standard deviation and the calculated Doppler limit for each mode.  

\SuppDoppAll

\section{ac quadratic Zeeman shift}
To resolve directional ambiguity, we measure $\langle B_{AC}^2\rangle$ with the magnetic field and cooling oriented in three different conditions.  We label these as $Q$ (original quantization axis, data in first panel of main text Fig. 4A), $H$ (horizontal, second panel), and V$^\prime$ ($\sim10\degree$ from vertical, third panel).  The quantization axes $Q$ and $H$ are labeled in Fig. \ref{fig:SuppRaman}, while the axis $V$ is out of plane.  In each of the three quantization field configurations, the hyperfine frequency shift due to the ac magnetic field $\langle B_{AC}^2\rangle$ is measured using interleaved spectroscopy of two $^{25}\mathrm{Mg}^+$ hyperfine transitions via the method presented in \cite{BrewerMagconst2019}.  For each quantization direction, we repeat this measurement at several trap RF powers, yielding the data shown in Fig. 4A.  

The measurements in any direction are not independent from the others; a measurement along a quantization axis $\hat{z}$ gives a frequency shift proportional to $\langle B_z^2\rangle+\frac{1}{2}\langle B_\perp^2\rangle$, where $B_\perp$ is the ac magnetic field perpendicular to $\hat{z}$.  Additionally, the ``vertical'' quantization axis is actually offset by $\sim10\degree$ to match the propagation direction of the cooling and detection beam.  We therefore simultaneously fit all the data to extract the magnetic field along each quantization axis.  

We label the ac-induced frequency shift (i.e. the frequency shift on $m_F=0\rightarrow 0$ after subtracting the effect of the dc field, as discussed in the main text) as $\Delta_{00}\vert_{Q,H,V^\prime}$.  The sum of squared secular frequencies, $F^2$, is proportional to the RF power applied to the trap electrodes \cite{WubbenaCooling2012}.  We fit to the system of equations,
\begin{equation}
\begin{gathered}
\Delta_{00}\vert_{Q}=\omega_0+\left(M_Q^2+\frac{1}{2}M_H^2+\frac{1}{2}M_V^2\right)F^2,\\
\Delta_{00}\vert_{H}=\omega_0+\left(\frac{1}{2}M_Q^2+M_H^2+\frac{1}{2}M_V^2\right)F^2,\\
\Delta_{00}\vert_{V^\prime}=\omega_0+\left|
\begin{pmatrix}
    \frac{\cos\theta+1}{{2}\sqrt{2}}&\frac{\cos\theta-1}{{2}\sqrt{2}}&-\frac{\sin\theta}{{2}}\\
    \frac{\cos\theta-1}{{2}\sqrt{2}}&\frac{\cos\theta+1}{{2}\sqrt{2}}&-\frac{\sin\theta}{{2}}\\
    \frac{\sin\theta}{\sqrt{2}}&\frac{\sin\theta}{\sqrt{2}}&\cos\theta
\end{pmatrix}
\begin{pmatrix}
    M_Q\\M_H\\M_V
\end{pmatrix}
\right|^2F^2,
\end{gathered}
\end{equation}
where we have accounted for for the angle $\theta\approx 10\degree$ between V$^\prime$ and the trap vertical axis V.  $M^2=d(\omega_{0\rightarrow0})/dF^2$ are the frequency shifts per trap RF due to ac magnetic fields along the trap coordinate axes $Q$, $H$, and $V$, and $\omega_0^2$ is the zero-field hyperfine splitting.  From the fit results $M$ and measured $F^2$, together with known $^{25}\mathrm{Mg}^+$ magnetic constants, we obtain the total squared ac magnetic field $\langle B_{AC}^2\rangle=0.85\pm0.09$ $\mu \mathrm{T}^2$.  We also obtain $\omega_0$, which gives a zero-field hyperfine splitting of $1,788,762,753.74\pm 0.13$ Hz.  

To quantify the uncertainty in the extracted values, we use a parametric bootstrapping method \cite{EfronBootstrap1979} to generate 10,000 data sets.  In addition to resampling each data point based on its statistical uncertainty, we also resample the angle $\theta$ from a distribution centered around $10\degree$.  We use a standard deviation of $3\degree$ constrained by trap geometry, but find that even a $5\degree$ standard deviation does not significantly increase the uncertainty on $\langle B_{AC}^2\rangle$.  

We note that the measured hyperfine splitting differs by 0.89 Hz from that reported in \cite{BrewerMagconst2019}.  We have identified the discrepancy as arising from a direct digital synthesizer control issue in the previous measurement which shifted the microwave drive frequencies sent to the ion by $3.5\times10^9/2^{32}=0.82$ Hz, with the remaining 0.07 Hz discrepancy being within the combined measurement uncertainty. This error does not affect the systematic evaluation of the previous clock and will be addressed in an Erratum.

\section{Cooling light Stark shift}

To constrain the Stark shift due to the continuous Doppler cooling beam, we directly measure the polarizability of the $^{27}$Al$^+$ clock transition by 280 nm light.  We make this measurement by applying a strong 280 nm laser, which is otherwise used as half of a pair driving $^{25}$Mg$^+$ Raman transitions.  We measure the Stark shift this beam causes on the clock transition, as well as its saturation parameter S, to get the polarizability in units of $\Delta\nu/S$.  Finally, to constrain the Stark shift due to the cooling laser, we regularly measure its saturation parameter during clock operation.

\SuppRaman

The apparatus has $^{25}$Mg$^+$ Raman beam pairs with four different orientations, shown in figure \ref{fig:SuppRaman}.  All share the same red-detuned Raman beam.  The four possible blue-detuned beam paths make Raman pairs with different k-vectors: nearly zero for the ``co-propagating'' pair, along the trap axis for the ``axial'' pair, in-plane orthogonal to the trap axis for the ``radial'' pair, or a linear combination of all three directions for the ``vertical'' pair.  The co-propagating beams are tightly focused onto the $^{25}$Mg$^+$ ion and used to check the orientation of the $^{25}$Mg$^+$-$^{27}$Al$^+$ crystal, while the other three Raman beams and the cooling/detection beam are less tightly focused, such that the beam waists are large compared to the inter-ion separation.  We use the ``radial'' blue-detuned Raman beam as the Stark shifting beam.  This beam has a maximum saturation parameter more than 500 times larger than we typically use for Doppler cooling, resulting in a shift as large as $-2.8$ Hz (or fractionally, $-2.5\times 10^{-15}$) on the $^{27}$Al$^+$ clock transition. 

Before the Stark shift measurement, we calibrate the saturation parameter of the beam by measuring Raman transition Rabi rates.  In the limit of large detuning from a single intermediate state, the Rabi rate for a Raman transition is given by 
\begin{equation}
\label{eqn:ham1}
    \Omega=g_R g_B \frac{\Omega_R^0\Omega_B^0}{2\Delta}, 
\end{equation}
where $\Omega_{R,B}^0=\Gamma \sqrt{\frac{S_{R,B}}{2}}$, $g_{R,B}$ are the Clebsch-Gordan coefficients for the red and blue transitions to the intermediate state, $\Delta$ is the common-mode Raman detuning, $\Gamma$ is the decay rate for the intermediate state, and $S_{R,B}=\frac{I_{R,B}}{I_S}$ are the saturation parameters for the red and blue beams in terms of the saturation intensity of the transition.  

We first measure the Rabi rate of the carrier transition driven by the co-propagating Raman beams.  These beams are overlapped in a single-mode UV-transmitting optical fiber then polarized using a Glan-laser polarizer.  We measure the intensity ratio $R_{RB}$ between these beams using a photodiode after a 90:10 beam splitter, and can write $S_B=S_R R_{RB}$.  (We check the linearity of this photodiode in the relevant power range by measuring the co-carrier Rabi rate at different beam powers and find $<1 \%$ deviation, contributing minimally to the measurement uncertainty.)  Letting the Clebsch-Gordan coefficients for the red and blue co-propagating beams be $g_R$ and $g_B$, we write the saturation parameter $S_R$ of the red beam in terms of the measured Rabi rate for the co-carrier transition $\Omega_{co}^{meas}$ as
\begin{equation}
\label{eqn:ham2}
S_{R}=\frac{4\Delta\Omega_{co}^{meas}}{\Gamma^2 g_R g_B\sqrt{R_{RB}}}.
\end{equation}

Next, we measure the Rabi rate of the ``radial'' beam pair, $\Omega_{rad}^{meas}$.  Labeling the ``radial'' blue-detuned Raman beam's Clebsch-Gordan coefficient and saturation parameter as $g_{rad}$ and $S_{rad}$, and inserting the above expression for $S_R$, we can extract the saturation parameter
\begin{equation}
\label{eqn:ham3}
S_{rad}=\frac{4\Delta g_B \sqrt{R_{RB}}}{\Gamma^2 g_R g_{rad}^2}\frac{(\Omega_{rad}^{meas})^2}{\Omega_{co}^{meas}}.
\end{equation}

This gives an absolute calibration of the Stark shifting Raman beam intensity.  Next, we invert the ion crystal order, such that the $^{27}$Al$^+$ ion sees exactly the calibrated beam.  To measure the Stark shift from this beam on the clock transition, we use an interleaved self-comparison: we alternate between locking the clock laser frequency to the ion with the Stark shifting beam off, and measuring the average frequency shift with the Stark shifting beam on.  The frequency shift divided by the saturation parameter gives the differential polarizability of the $^{27}$Al$^+$ clock transition at 280 nm.  

We repeat this measurement at several different Stark shifting beam powers; each measurement is an independent calibration of the polarizability, and their agreement is a check against nonlinearity or other possible systematic effects.  The results are presented in Fig. 4 of the main text.  Including systematic uncertainty from atomic structure coefficients, as well as statistical uncertainty from the measurements, the fractional frequency shift and uncertainty on the clock transition are $\Delta\nu/\nu=-(2.59\pm0.06)\times10^{-17}\times S$.

To know the Stark shift on the clock transition from the continuous Doppler cooling light, we also need to know the saturation parameter during cooling.  We measure this using a depumping experiment similar to \cite{ChouComparison2010}.  We prepare the $^{25}$Mg$^+$ ion in the ``dark'' qubit state, $\lvert{}^2$S$_{1/2}, $F$=2, \mathrm{m}_\mathrm{F}$=$-2\rangle$, and measure the rate at which the Doppler cooling light off-resonantly depumps it to the ``bright'' state $\lvert{}^2$S$_{1/2}, $F$=3, \mathrm{m}_\mathrm{F}$=$-3\rangle$.  After a depumping time $t$, the ion fluorescence count rate is
${C=d+b\left(1-e^{-S\frac{t}{\tau}}\right)}$,
where we have labeled the count rate from the bright state as $b$ and the dark counts as $d$ (e.g. scattered photons from the detection beam on the vacuum chamber, etc).  $\tau=213.5\pm 0.4$ $\mu$s is the depumping time constant for S=1, calculated from $^{25}$Mg$^+$ atomic structure coefficients and our Doppler cooling beam detuning.  We integrate over the detection time $t_d$ to account for continued depumping by the detection beam, giving the total counts ${C_{tot}=(b+d)t_d+\tau_0\frac{b}{S}\left(e^{-S\frac{t+t_d}{\tau}}-e^{-S\frac{t}{\tau}}\right)}$. 

Scanning $t$ and fitting the resulting data gives the saturation parameter $S$.  To control for drifts in cooling laser power during clock operation, we interleave measurements of $S$ with a low duty cycle ($\sim1\%$).  We average these measurements after each day of clock operation; across many such days, the average saturation parameter is $0.148$ and the typical uncertainty is $\sim\pm 0.004$.  

Finally, while the Doppler cooling beam waist is large compared to the separation between ions, we still must account for any possible difference in laser power between the $^{25}$Mg$^+$ and $^{27}$Al$^+$ ion positions.  Before and after operating the clock, we measure the ratio of saturation parameters on $^{25}$Mg$^+$ when it is in its normal position and with the ion crystal orientation reversed.  We scale the saturation parameter obtained from interleaved measurements by this ratio, and add the ratio uncertainty in quadrature.  We then apply a correction to the day's clock data. 
 The average shift on the clock transition from this effect is $\Delta\nu/\nu=-(37.2\pm 2.0) \times 10^{-19}$, with the exact shift and uncertainty depending on the measurements on a given day.

\section{Excess micromotion}
A previous version of the ion trap used here exhibited excess micromotion (EMM) that could not be compensated by adjusting the dc shim voltages, nor by adding rf potentials to the compensation voltages.  This motivated the operation of the clock at a lower trap drive frequency and correspondingly lower secular motion frequencies, to reduce the overall micromotion shift~\cite{BrewerAccuracy2019}.  EMM ultimately limited the accuracy of the clock due to uncertainty in measurements of the residual time-dilation shift.  The source of the uncompensatable micromotion was found to be a phase shift between the rf voltages on opposite electrodes coming from a combination of unbalanced capacitances and resistance in the gold traces on the trap wafer.  To improve this, the trap was redesigned based on a thicker diamond wafer (500 $\mathrm{\mu}$m rather than 300 $\mathrm{\mu}$m) and rerouted gold traces to balance the capacitances of the traces going to opposite rf electrodes.  The sputtered gold traces were made thicker (3 $\mathrm{\mu}$m rather than 500 nm) to reduce resistance.  The result is reduced EMM due to phase shifts at the electrodes that, if necessary, can further be compensated by applying rf voltages to the compensation electrodes and endcaps. 

We characterize EMM using the resolved-sideband method \cite{KellerMMmeasurement2015}.  To measure in three dimensions, we use the ``axial'' and ``radial'' Mg$^+$ Raman beam pairs (see Fig. \ref{fig:SuppRaman}) and either the ``vertical'' Mg$^+$ Raman pair or Al$^+$ $^3\mathrm{P}_1$.  The ``vertical'' beams are not fully orthogonal to the other two - by 10$\degree$ for $^3\mathrm{P}_1$ or $\sim45\degree$ for ``vertical'' Raman k-vector; to account for this non-orthogonality, we use a Monte Carlo approach which has been described previously \cite{BrewerAccuracy2019}.  This Monte Carlo analysis also incorporates statistical uncertainty in the sideband ratio measurements, uncertainty in the directions of each laser beam, and the unknown relative phase between the EMM components.  

\MMSupp

In some cases, the Rabi frequency of a well-compensated micromotion sideband is longer than the coherence time of the lasers used to measure it.  Figure \ref{fig:MMSupp} shows two micromotion sideband measurements.  In panel A, the micromotion sideband pi-time can be clearly resolved, whereas in panel B the coherence time is insufficient to reliably determine it.  In these cases, we can at best estimate a bound on the sideband Rabi rate, and therefore an upper bound on the EMM along the relevant beam axis.  To do this, we compare chi-squared values for fits to the data with the sideband pi-time fixed to values between 0.1 ms and 4 ms.  As a lower bound, we use the pi-time for which the fit chi-squared is double that for a simple exponential decay.  Pi-times faster than this are increasingly inconsistent with the data, and ruled out by the lack of ``Rabi flopping''.  As input to the Monte Carlo in these cases, we use a flat distribution with equal probability for values between 0 and the EMM bound.

Figure \ref{fig:MMSupp}C shows a set of EMM measurements taken before and after operating the clock for periods between 4 and 7 hours.  The compensation voltages and ion position are stable enough that an active micromotion servo is unnecessary.  For the systematic shift and uncertainty contributions from EMM, we use the weighted average of all measurements and an error bar that includes zero.  This represents our model uncertainty about the origin of the decoherence that prevents us from accurately measuring small EMM values; the covered range includes all of the measured values.  This gives a shift and uncertainty of $\Delta\nu/\nu=-(1.6 \pm 1.6)\times 10^{-19}$.

\section{Vacuum and collision shift}

The vacuum system was constructed to achieve low pressures and enable operation for long probe times and potentially with multiple ions.  Significant design features, noted in Fig. \ref{fig:SuppVacuum}A, include a large main chamber and large diameter connections to pumps; construction of all vacuum elements except pumps from titanium, to minimize hydrogen outgassing; and inclusion of two non-evaporable getter pumps in the main chamber, with direct line-of-sight and high conductance to the ion trap.  We evaluate vacuum quality by measuring the reorder rate of a two-ion crystal, using a tightly focused Raman beam.  Example data is shown in Fig. \ref{fig:SuppVacuum}B.  During a total 79 hours of measurement, we observed 25 reorder events, for a reorder period of 11,376 seconds.

Following the procedure described in \cite{HankinBackgroundgas2019}, we calculate a reorder potential energy barrier of 3.6 $\mathrm{K} \times k_B$ and a background gas pressure of $(2.5 \pm 1.3) \times 10^{-10}$ Pa $[(1.8\pm0.9)\times10^{-12}$ Torr].  This vacuum pressure is more than 150 times lower than that in the previous-generation NIST aluminum ion clock.  Therefore, even taking a worst-case bound for the phase shift and Doppler shift components, we estimate a shift $\Delta\nu/\nu=-(0.26\pm0.37)\times10^{-19}$ by scaling from the previous result and accounting for Doppler cooling during the probe time.\\

\SuppVacuum

\FloatBarrier 

\clearpage

\end{document}